\documentclass[12pt,a4paper]{article}
\usepackage{amsmath,amsthm,amssymb,natbib}
\usepackage{graphicx}
\usepackage{enumerate}
\usepackage{booktabs}
\usepackage{MnSymbol}

\newtheorem{definition}{Definition}
\newtheorem{theorem}[definition]{Theorem}
\newtheorem{lemma}[definition]{Lemma}
\newtheorem{corollary}[definition]{Corollary}
\newtheorem{proposition}[definition]{Proposition}
\newtheorem{example}{Example}
\newtheorem{alg}{Algorithm}

\begin{document}
\title{\bf Iterative conditional replacement algorithm for conditionally specified models}
\author{Kun-Lin Kuo\\
    Institute of Statistics, National University of Kaohsiung, Kaohsiung, Taiwan\\
    and \\
    Yuchung J. Wang\thanks{Corresponding author: yuwang@camden.rutgers.edu}\\
    Department of Mathematical Sciences, Rutgers University, Camden, NJ, USA}
\date{}
\maketitle

\begin{abstract}

The sample-based Gibbs sampler has been the dominant method for approximating joint distribution from a collection of compatible full-conditional distributions.  However for conditionally specified model, mixtures of incompatible full and non-full conditional distributions are the realities; but, their updating orders are hard to identified. We propose a new algorithm, the Iterative Conditional Replacement (ICR), that produces distributional approximations toward the stationary distributions, dispensing Markov chain entirely.   ICR always converges, and it produces mutually stationary distributions, which will be consistent among one another when the conditional distributions are compatible. Examples show ICR to be superior in quality, while being more parallelizable and requiring little effort in monitoring its convergence.
Last, we propose an ensemble approach to decide the final model.
\end{abstract}

\noindent%
{\bf Keywords}: Dependency network; $I$-projection;
Method of alternating projection;  Mutually stationary distributions; Unsupervised leaning.
\vfill

\section{Introduction}
Using the two cultures of \citet{Breiman2001}, the assumption of a joint distribution is  data modeling, whereas conditionally specified model (CSM)---specifying a joint distribution via conditional distributions---belongs to the camp of algorithmic modeling.   %The journal \emph{Observational Studies}~(2021) devoted an entire issue discussing the differences/convergence of the two approaches.
A typical example is in multiple imputation: explicit full multivariate (Bayesian) models versus MICE \citep[multiple imputation by chained equations,][]{Raghunathan2001,vanBuuren2007}.  However,  Markov random field \citep{Kaiser2000}, spatial modeling \citep{Besag1974}, and dependency networks \citep{Heckerman2000} had been shown that the conditional approach offers certain advantages. CSM can be used to compose joint models from data collected over spatial ranges or temporal stages, because
it would be unrealistic to simultaneously articulate a joint model for  a large number of variables.
A better is to locally model a small number of variables, then combine those submodels into a joint model, like embedding pieces of a jigsaw puzzle into a complete picture.
%That is, the nodes are grouped into subsets, and each subset is  modeled either marginally or conditionally.
Our algorithm will make the process of modeling locally and synthesizing  globally easier.
Formally, CSM determines a joint distribution for $\mathbb{X}=(x_1,\ldots,x_d)$ after three stages of maneuvers:

\begin{enumerate}[\sf {Stage} I.]
\item Conditional modeling: Built a predictive conditional model from data for every $x_i \equiv \{i\}$ using a subset of $\mathbb{X} \backslash \{x_i\}\equiv \{-i\}$ as the predictors via a regularized modeling or machine learning algorithm, such as regression, classification, or a neural network.  Let the learning outcome be $\{f_{i|c_i}: 1 \le i \le d\}$,  where $c_i \subseteq \{-i\}$.  Or more directly, a conditional model, $\{ f_{a_i|b_i}: 1 \le i \le L \}$, has already been formulated by domain experts using subject matter knowledge and algorithms of her choice, where $a_i$ and $b_i$ are non-intersecting subsets of $\mathbb{X}$.
    For spatial data, $c_i$ ($b_i$) is commonly known as the ``neighbors'' of $x_i$ ($a_i$); in general, $c_i$ ($b_i$) is the covariates used to predict $x_i$ ($a_i$).
\item Synthesize (from local to global): Embed the conditional distributions, $\{f_{i|c_i}: 1 \le i \le d\}$ or $\{f_{a_i|b_i}: 1 \le i \le L \}$, into joint distributions of $\mathbb{X}$. Nodes of $\mathbb{X}$ may be divided into groups.   Within each group, the synthesis  produces intermediate distribution.  These intermediate distributions then propagate in phases to the entire $\mathbb{X}$, with the sequential orders of propagation playing  a critical role.  %Therefore,
    %propagation is a useful perspective to comprehend the step-by-step formation of joint distributions
\item Optimize: Different sequences to propagate the intermediate distributions may result in different joint distributions.  The entire collection of  stationary joint distributions, produced in {\sf Stage II}, make up an ensemble, and it is the ensemble that makes the final model of $\mathbb{X}$.

\end{enumerate}
The final outcome of a CSM will depend on both the data and the algorithms used in the three stages.  Here, we propose an algorithm to divide and to synthesize, and recommend another algorithm for the optimization attendant to {\sf Stage III}. Absent the concerns of {\sf Stages II} and {\sf III}, much algorithmic creativity remains available in {\sf Stage I}.

A conditional model of {\sf Stage I} is said to be compatible if  a joint distribution exists, from which  every conditional or marginal distribution can be derived.  In such a circumstance, the output of a synthesis should be unique.  Moreover, a CSM is said to be sufficient if it has enough information to identify a joint distribution of $\mathbb{X}$.
A conditional distribution involving all the variables in $\mathbb{X}$ is called  a full-conditional and is expressed as $f_{i|-i}$ or $f_{a_i|-a_i}$;
otherwise, it is a non-full conditional: $f_{a_i|b_i}$, $a_i\cup b_i \ne \mathbb{X}$. When the CSM is $\{f_{i|-i}: 1 \le i \le d\}$ and the Gibbs sampler (GS) is used for synthesis, there can be up to $d!$ (systematic scan) stationary distributions, one for each permutation of $(1,\ldots,d)$ \citep{Chen2015}.

Most CSM papers only consider full-conditional models that mimic  the Bayesian computation \citep{Smith1993}.  However, proposing a full-conditional for every variable of $\mathbb{X}$ is impractical; %, if not more, as proposing one joint model;
in stead, a mixture of full and non-full conditionals is  a more realistic approach.   Therefore, practical  synthesis must
be able to accommodate combinations of full and non-full conditionals.
\citet{vanDyk2008} invented partially collapsed Gibbs sampler (PCGS): the GS based on combinations of \emph{compatible} full and non-full conditionals.
They discovered that PCGS must follow specific updating orders to draw correct samples.
Another difference between Bayesian computation and CSM is that  approximating the posterior distribution  is not the main objective of GS,  while joint distribution of $\mathbb{X}$ is the only focus of CSM.  Here, we invented the Iterative Conditional Replacement algorithm (ICR) which produces distributions, not samples.  ICR will simultaneously compute several joints and/or marginal distributions
\emph{regardless} of  compatibility and its convergence is guaranteed.
When the CSM is compatible, ICR will approximate the unique stationary distribution; otherwise, the joint distributions would be many and different.  More critically, we devise simple rules to identify all the permissible updating orders.% for all CSMs. %mixtures of full and non-full conditionals.
The examples below show that ICR is computationally more robust and flexible than sample-based methods.

Traditionally, compatibility must be confirmed before GS or PCGS sampling can start; otherwise, the Markov chains can become null.
In contrast, ICR  cycles through a permissible updating order, and produces mutually stationary  distributions.
%These distributions will be used as input to find an optimal joint using the ensemble method proposed by \citet{Chen2013}.
Moreover, there are compatible and sufficient CSM, such as $\{ f_{1|23}, f_{2|13}, f_3\}$, that PCGS cannot sample, because it cannot pass the dependence of $(x_1,x_2)$ back to $x_3$. We propose ``divide-then-ICR'' strategy: first, the CSM is divided into suitable groups such that permissible updating orders within each group can be found;  second, apply ICR to each group and produce  (intermediate) distributions for subsets of $\mathbb{X}$. Finally, use ICR again to combine intermediate distributions into joint distributions or marginal distributions.  For example,  $\{ f_{1|23} ,f_{2|13}, f_3\}$ is first divided into $\{ f_{1|23}, f_{2|13} \}$ and $\{ f_3\}$.  From  $\{ f_{1|23}, f_{2|13} \}$, ICR computes  two stationary $\pi_{12|3}^{(1,2)}$ and $\pi_{12|3}^{(2,1)}$, where the superscripts indicate different updating orders. We multiply either distribution by $f_3$ and get the two mutually stationary joint distributions: $\pi_{123}^{(1,2)}$ and  $\pi_{123}^{(2,1)}$.  If these two joints are equal, the original CSM is deemed compatible.  The {\sf Stage III} optimization is to find a mixture, $\alpha\pi_{123}^{(1,2)}+ (1-\alpha)\pi_{123}^{(2,1)}$, that minimizes the deviance relative to the original CSM.

In the past, there have been many algebraic proposals to verify the compatibility among full conditionals, for example, \citet{Wang2008} and \citet{Arnold2002}.
However, how to verify the compatibility between full and non-full conditionals is still very much an open problem.
Here is a case that computations can answer  algebraically difficult question;
we prove that the CSM is compatible when the multiple stationary distributions computed by ICR are the same.
In the examples below, benefits of ICR are highlighted by its capacity to handle (a) incompatible CSM; (b) reducible CSM whose support is partitioned; (c) the conditional density is sticky for GS to sample (slow mixing); and (d) the CSM that divide-then-ICR can synthesize, whereas PCGS cannot.

ICR is introduced in Section~\ref{sec:2}, first for full conditionals, then for combinations of full and non-full conditionals.  ICR is cyclically doing $I$-projections  among spaces defined individually  by  each conditional distribution.
%A new notion of stationarity is discussed.
Examples are in Section~\ref{sec:3}. Many times, ICR cannot be applied to a CSM directly;  but
partitioning a CSM into several smaller CSM enables ICR to be applied locally. %In Section~\ref{sec:3-1}, we show how divide-then-ICR works.
%ICR is a new algorithm for CSM; its
Historical connections of ICR with other algorithms,
such as GS, power method, and alternating projection
are addressed in Section~\ref{sec:4}. Section~\ref{sec:5} contains a brief conclusion.

\section{The iterative conditional replacement algorithm}\label{sec:2}
Hereafter, conditional and marginal distributions/densities will be abbreviated as conditional(s) and marginal(s).
A joint density is denoted by $p$, $q$, $\pi$, $f$, or $g$ without subscript, while their marginal and conditional densities have subscripts and are denoted as
$\pi_1$, $p_{ij}$, $q_a$, $q_{-a}$, $f_{i|-i}$, $g_{12|34}$,
where $1=\{x_1\}$, $ij=\{x_i,x_j\}$, $a=\{x_i: i\in a\}$, $-a=\{x_i: i \not \in a\}$, $i|-i=\{x_i|x_j, j\ne i\}$, and $12|34 =\{x_1,x_2|x_3,x_4\}$.
We also reserve $f_{a_i|b_i}$ and $g_{a_j|b_j}$ for the conditional distributions in a CSM,
$p$ and $q$ as the distributions produced during ICR iterations,
and $\pi^{(i_1,\ldots,i_d)}$ for the stationary joint distribution updated in the order of $(i_1,\ldots,i_d)$.
Moreover, let ${\cal S}(f)$ and ${\cal S}(f_{i|-i})$ be the support of $f$ and $f_{i|-i}$, respectively; ${\cal S}(q_a)$ be the support of $q_a$.  We always assume ${\cal S}(f_{j|-j}) = {\cal S}(f_{i|-i})$ for all $(i,j)$.  A $d$-dimensional joint density $f$ is said to satisfy the total positivity condition if ${\cal S}(f)= {\cal S}(f_1)\times \cdots\times {\cal S}(f_d)$.
We use Kullback-Leibler divergence, called K-L divergence hereafter, as the measure of deviance that drives ICR's search.  The K-L divergence is defined as
\[ I(p;q)=\sum_x p(x) \log \frac{p(x)}{q(x)}.\]

\subsection{ICR for conditionally specified models of full conditionals}\label{sec:2-1}
Let the CSM be $\{f_{j|-j}: 1 \le j \le d\}$, and ${(i_1 ,i_2, \ldots, i_d)}$ and ${(i_2, \ldots, i_d, i_1)}$ be two adjacent updating orders.  \citet{Kuo2019} prove the following properties for $\{\pi^{(i_1,\ldots,i_d)} \}$:
%, where each
%$\pi^{(i_1,\ldots,i_d)} $ is the stationary distribution of  the Gibbs sampler,  updating in the order of $(i_1, \ldots,i_d)$:

%Let $(i_1,\ldots,i_d)$ be a permutation of $(1,\ldots,d)$, and let $\{\pi^{(i_1,\ldots,i_d)} \}$ denote the
%stationary distributions of $\{f_{j|-j}: 1 \le j \le d\}$ updating in the order of $(i_1, \ldots,i_d)$.
%The following was proved in \citet{Kuo2019}:
\begin{enumerate}[(H1)]
\item Stationary distributions $\pi^{(i_1 ,i_2, \ldots, i_d)}$ and $\pi^{(i_2, \ldots, i_d, i_1)}$, respectively, have $f_{i_d|-i_d}$ and $f_{i_1|-i_1}$ as their conditionals;
\item $\pi^{(i_1, i_2, \ldots, i_d)}_{-i_1} = \pi^{(i_2, \ldots, i_d, i_1)}_{-i_1}$; and
\item $\pi^{(i_1, i_2, \ldots, i_d)}_{i_1} = \pi^{(i_2, \ldots, i_d, i_1)}_{i_1}$.
\end{enumerate}
Therefore, the goal of the algorithm is to formulate  sequences of joint distributions that monotonically approximate the  $\{ \pi^{(i_1, \ldots,i_d)} \}$ such that they collectively fulfill (H1)--(H3).  Requirements (H2) and (H3) are necessary for balancing the degrees of freedom between the CSM and the collection of all the stationary distributions.

To illustrate, consider a simple CSM ${\cal A}=\{f_{1|2},f_{2|1}\}$, and
define ${\cal C}_1=\{f_{1|2}\omega_2\}$ and ${\cal C}_2=\{f_{2|1}\nu_1\}$,
where $\omega_2$ and $\nu_1$ are  marginal densities of $x_2$ and $x_1$,
respectively. Let $q$ be a joint density having the same support of $f_{1|2}$.
The K-L divergence between $q$ and a $\tau= f_{1|2} \tau_2 \in {\cal C}_1$ satisfies the Pythagoras equality:
\[I(q;\tau)=I(q; f_{1|2}q_2) + I(f_{1|2}q_2;\tau),\]
which is proved in Appendix~\ref{app1}.
By choosing $\tau_2=q_2$, $I(f_{1|2}q_2;\tau)=0$ and minimization of $I(q;\tau)$ is achieved.  Thus, $I$-projection of $q=q_{1|2} q_2$ onto ${\cal C}_1$, is $f_{1|2} q_2$, so it is named conditional replacement.
By the same token, the $I$-projection of $q=q_{2|1} q_1$ onto ${\cal C}_2$ is $f_{2|1}q_1$.
Let the iterations begin from a $q^{(0)} $.
The following  alternating $I$-projections between $ {\cal C}_1$ and $ {\cal C}_2$ produce  two sequences of joints:
\[
q^{(2k+1)}=f_{1|2}q^{(2k)}_2 \in {\cal C}_1 \mbox{ and } q^{(2k+2)}=f_{2|1} q^{(2k+1)}_1 \in {\cal C}_2, \mbox{ with $k=0,1,2,\ldots$.}
\]
Throughout, (H1) holds for both $\{ q^{(2k+1)} \}$ and $\{ q^{(2k+2)} \}$.
The choices of $q^{(2k+1)}_2= q^{(2k)}_2$ and  $q^{(2k+2)}_1= q^{(2k+1)}_1$ not only  minimize the K-L divergence, but also satisfy (H2).
Next, (H3) provides the metric to detect the convergence of ICR;
$I$-projections will be stopped at $t$ when $q^{(2t+1)}_1=q^{(2t)}_1$ and $q^{(2t+2)}_2=q^{(2t+1)}_2$.
Numerically, stop ICR at $t$-th iteration when $M(t)=I(q_1^{(2t)};q_1^{(2t+1)})+I(q_2^{(2t+1)};q_2^{(2t+2)}) < 10^{-10}$.
Upon convergence, we designate $q^{(2t+1)}$ as $\pi^{(2,1)} \in {\cal C}_1$ and $q^{(2t+2)} $ as $\pi^{(1,2)} \in {\cal C}_2$.  % The marginal matchings: $\pi^{(1,2)}_i= \pi^{(2,1)}_i, i=1,2 $  reduce the degrees of freedom of   $\{\pi^{(1,2)},~ \pi^{(2,1)} \}$ by $r+c-2$,  thus match the degrees of freedom of $\{ f_{1|2}, f_{2|1} \}$, where $r$ and $c$ are the degrees of freedom of $x_1$ and $x_2$, respectively.

The following proposition follows from Theorem~\ref{thm:convergence} to be proved later.
\begin{proposition}
Both $I(\pi^{(2,1)}; q^{(2k+1)})$  and $I(\pi^{(1,2)};q^{(2k+2)})$ decrease
to $0$ as $k\to \infty$.
\end{proposition}
Due to the total variation norm inequality, $\|P-Q\| \leq \sqrt{\frac{1}{2}~I(P;Q)} $,
$\|q^{(2k)} - \pi^{(1,2)}\| \rightarrow 0$ and $\|q^{(2k+1)} - \pi^{(2,1)}\| \rightarrow 0$.

\begin{proposition}
$\pi^{(1,2)} =\pi^{(2,1)}$ if and only if $\{f_{1|2}, f_{2|1}\}$ are compatible.
\end{proposition}
\begin{proof}
$\pi^{(1,2)} =\pi^{(2,1)}$ implies ${\cal C}_1 \cap {\cal C}_2 \ne \emptyset$, thus compatible.  When $\{f_{1|2}, f_{2|1}\}$ are compatible if and only if they have the same odds ratios.
Two distributions are the same if and only if they have the same odds ratios and the same marginal densities, which ICR is designed to achieve, i.e., (H2) and (H3).
\end{proof}

\citet{Wang2008} has an algebraic check of the compatibility between $f_{1|2345}$ and $f_{2|1345}$ without iteration.
Alternatively, ICR  begins with an arbitrary $q^{(0)}_{2|345}$  and computes $q^{(2k+1)}_{12|345}=f_{1|2345}q^{(2k)}_{2|345}$ and $q^{(2k+2)}_{12|345}=f_{2|1345} q^{(2k+1)}_{1|345}$, until they converge to $\pi^{(1,2)}_{12|345}$ and $ \pi^{(2,1)}_{12|345}$, respectively.   Regardless of the initial $q^{(0)}_{2|345}$ , $\pi^{(1,2)}_{12|345} =\pi^{(2,1)}_{12|345}$ confirms compatibility.

For $d=3$ and CSM: $\{f_{1|23}, f_{2|13}, f_{3|12}\}$, define
${\cal C}_i=\{f_{i|-i} v_{-i}\}$ for $i=1,2,3$, where $v_{-i}$ is any marginal density of $x_{-i}$.
There are two updating orders: clockwise: ${\cal C}_1\to {\cal C}_2 \to {\cal C}_3 \to {\cal C}_1 \to \cdots$; and counter-clockwise: ${\cal C}_1\to {\cal C}_3 \to {\cal C}_2 \to {\cal C}_1 \to \cdots$.
The three stationary distributions of clockwise sequence are
$ \pi^{(1,2,3)} \in {\cal C}_3,$ $\pi^{(2,3,1)} \in {\cal C}_1$ and $\pi^{(3,1,2)} \in {\cal C}_2$, and they are called circularly-related, and ICR approximates them with the following iterations:
%The ICR for the clockwise sequence is as follows:
\[ q^{(3k+1)}=f_{1|23} q^{(3k)}_{23}, q^{(3k+2)}=f_{2|13} q^{(3k+1)}_{13} \mbox{ and }
 q^{(3k+3)}=f_{3|12} q^{(3k+2)}_{12},k=0,1,2,\ldots .\]
The above marginalization-then-multiplications is designed to satisfy both (H1) and (H2). And
ICR stops iterations when (H3): $q^{(3t)}_{1}=q^{(3t+1)}_{1}$ $q^{(3t+1)}_{2}=q^{(3t+2)}_{2}$ and $q^{(3t+2)}_{3}=q^{(3t+3)}_{3}$, are reached.
Numerically,  ICR stops when  $M(t)=I(q_{1}^{(3t)};q_{1}^{(3t+1)})+I(q_{2}^{(3t+1)};q_{2}^{(3t+2)})
+I(q_{3}^{(3t+2)};q_{3}^{(3t+3)})<10^{-10}$.  The following proposition follows from Theorem~\ref{thm:convergence}.

\begin{proposition}
For the clockwise updating order, the three sequences of joint densities converge,
respectively, to their stationary distributions.  That is, as $k \rightarrow \infty$,
$q^{(3k+1)} \to \pi^{(2,3,1)} \in {\cal C}_1$,
$q^{(3k+2)} \to \pi^{(3,1,2)} \in {\cal C}_2$
and $q^{(3k+3)} \to \pi^{(1,2,3)} \in {\cal C}_3$ in K-L divergence.
\end{proposition}

\begin{proposition}
CSM: $\{f_{1|23}, f_{2|13}, f_{3|12}\}$ are compatible if and only if $\pi^{(1,2,3)} =\pi^{(2,3,1)}=\pi^{(3,1,2)}$.
\end{proposition}

Let ${\cal D}=\{1,\ldots,d\}$ represent $(x_1,\ldots,x_d)$.  Consider the conditional model:
${\cal A}=\{ f_{a_i|-a_i}: 1 \le i \le L\}$, with  $\bigcup_{i=1}^L a_i={\cal D}$.
Again define ${\cal C}_{a_i}=\{f_{a_i|-a_i} v_{-a_i} \}$, where $v_{-a_i}$ is any $x_{-a_i}$-marginal density.
For a fixed updating order: ${\cal C}_{a_1} \to {\cal C}_{a_2}\to \cdots \to {\cal C}_{a_L}$,  the $L$ circularly-related stationary distributions are
\[{\cal P}=\{\pi^{(a_2,\ldots,a_L,a_1)} \in {\cal C}_{a_1},
\pi^{(a_3,\ldots,a_L,a_1,a_2)} \in {\cal C}_{a_2}, \ldots, \pi^{(a_1,a_2,\ldots,a_L)} \in {\cal C}_{a_L}\}.\]
We start with $q^{(0)}= f_{a_L|-a_L} w_{-a_L} \in {\cal C}_{a_L}$.
One cycle of ICR consists of $L$ $I$-projections.  For $1 \le i \le L$ , the conditional replacements for (H1) and (H2) are:
\[q^{(Lk+i)}= f_{a_i|-a_i} q^{(Lk+i-1)}_{-a_i} \in {\cal C}_{a_i},\ k=0,1,\ldots.\]
The iterations stop at $t$ when $q^{(Lt+i)}_{a_i}=q^{(Lt+i-1)}_{a_i}$ for every $1\le i \le L$, that is, (H3).
Numerically, $\sum_{i=1}^{L} I(q_{a_i}^{(Lt+i)};q_{a_i}^{(Lt+i-1)}) < 10^{-10}$ is used to stop the iterations.

\begin{proposition}
If the $L$ stationary distributions
of ${\cal P}$ are the same, then the conditionals of ${\cal A}$ are compatible.
\end{proposition}
\begin{proof}
Because $\pi^{(a_{i+1},\ldots,a_{i-1},a_i)} \in {\cal C}_{a_i}$, the equalities of $L$ stationary distributions of ${\cal P}$ imply $\bigcap_{i=1}^{L}{\cal C}_{a_i} \ne \emptyset$, hence compatible.
\end{proof}

\subsection{ICR for unsaturated conditionally specified models (combinations of full and non-full conditionals)}\label{sec:2-2}
We shall name a CSM of exclusively full conditionals (Section~\ref{sec:2-1}), as a \emph{saturated} CSM, otherwise, the CSM is unsaturated. To model data, unsaturated CSM is more realistic.
But it is rarely discussed in the literature because the GS has a hard time sampling unsaturated CSM.
A major difficulty for GS  is finding the rules that identify the correct sequential orders to sample the non-full conditionals.
PCGS \citep{vanDyk2008}  is proposed to circumvent such issues, and  our algorithms will provide its theoretical justifications.
The following rules are quite intuitive from the perspective of conditional replacement.
Let an unsaturated CSM be represented by $\{f_{a_i|b_i}: 1 \le i \le L\}$ and  $\Delta=(\bigcup_{i=1}^{L} b_i) \backslash (\bigcup_{i=1}^{L} a_i)$.
Also, define ${\cal C}_{a_k}=\{f_{a_k|b_k} v_{b_k}=q_{a_k \cup b_k}\}$, where $v_{b_k}$ is a marginal distribution of $b_k$.

\begin{alg}\label{alg:1}\rm
Conditional replacement ($I$-projection) of  any $q_{a_i \cup b_i} \in {\cal C}_{a_i}$ onto ${\cal C}_{a_j}$ is permissible, written as ${\cal C}_{a_i} \rightharpoonup {\cal C}_{a_j}$, when the following two rules hold:
\begin{enumerate}[\sf Rule A.]
\item $b_j \subseteq a_i \cup b_i$.
\item $a_i \cap b_j \ne \emptyset$.
\end{enumerate}
\end{alg}

When  ${\cal C}_{a_i} \rightharpoonup {\cal C}_{a_j}$, we define the ICR mapping ${\mathbb P}: {\cal C}_{a_i} \to {\cal C}_{a_j}$ as ${\mathbb P}(q_{a_i \cup b_i})= f_{a_j|b_j}q_{b_j}$, where
%$q_{b_j}= \sum _{x \notin b_j} q_{a_i \cup b_i}( x \in a_i \cup b_i)$.
$q_{b_j}$ is the $x_{b_j}$-marginal density of $q_{a_i\cup b_i}$.
Marginalization of $q_{a_i \cup b_i}$ into $q_{b_j}$ can only be done when {\sf Rule A} holds.
Next, we consider applying ${\mathbb P}$ in cycle.
\begin{definition}
Let $(1^{\ast},\ldots, L^{\ast})$, be a permutation of $(1, \ldots,L)$ with $(L+1)^{\ast} \equiv 1^{\ast}$.  If every ${\mathbb P}$ mapping from  ${\cal C}_{a_{i^{\ast}}}$ to $ {\cal C}_{a_{(i+1)^{\ast}}}$ is permissible,
then $(a_{1^{\ast}},\ldots, a_{L^{\ast}})$ is said to be a permissible updating cycle for $\{f_{a_i|b_i}: 1 \le i \le L\}$, and is denoted as $\langle \langle a_{1^{\ast}},\ldots, a_{L^{\ast}} \rangle \rangle$.
\end{definition}

\begin{alg}[unconditioned ICR]\label{alg:2}\rm
Let the conditional model be ${\cal A}=\{ f_{a_i|b_i}: b_i \ne \emptyset, 1 \le i \le L\}$, $\Delta = \emptyset$, and $\bigcup_{i=1}^{L} a_i= \Lambda$.   When $\langle \langle a_{1^{\ast}},\ldots, a_{L^{\ast}}\rangle \rangle$,
%be a permutation of $(1, \ldots,L)$ with $(L+1)^{\ast} \equiv 1^{\ast}$.  If every $I$-projection from  ${\cal C}_{a_{i^{\ast}}}$ to $ {\cal C}_{a_{(i+1)^{\ast}}}$ is permissible,  then
ICR will synthesize joint and marginal distributions of ${\Lambda}$. %, and $\langle 1^{\ast},\ldots, L^{\ast}\rangle$ is referred to as a permissible updating cycle for ${\cal A}$.
In addition, the $I$-projections begin with a marginal distribution, $q^{(0)}_{b_{1^{\ast}}}$,  use $q^{(1)}=f_{a_{1^{\ast}}|b_{1^{\ast}}} q^{(0)}_{b_{1^{\ast}}}$ to initiate the iterations, and
${\mathbb P}^{k} (q^{(1)}) \in {\cal C}_{a_{r^{\ast}}}$ where $r=k\pmod L+1$.
\end{alg}

For example, CSM: $\{f_{12|3}, f_{4|123}, f_{3|124}, f_{5|1234}\}$
permits ${\cal C}_{12}\rightharpoonup {\cal C}_{4}\rightharpoonup {\cal C}_{3} \rightharpoonup {\cal C}_{5}$, but not ${\cal C}_{5} \rightharpoonup {\cal C}_{12}$ due to violation of Rule B; hence, Algorithm 2 cannot be applied.  Had  we  changed $f_{12|3}$ to $f_{12|35}$, then $\langle \langle 12,4,3,5\rangle \rangle$,% would permit  cyclic $I$-projections;
~and Algorithm~\ref{alg:2} will synthesize one joint, $\pi^*$, plus two  marginals:  $\{\pi_{1235}$, $\pi_{1234} \}$.
When $\pi^*_{1235}=\pi_{1235}$  and $\pi^*_{1234}=\pi_{1234}$,
the CSM is compatible.   In the following, we consider unsaturated CSM that specifies conditional distributions, not joints.   When $\Delta \ne \emptyset$, it can be shown that $\Delta \subset b_{i}$ for every $i$.

\begin{lemma}
Suppose that CSM
$\{f_{a_{i}|b_{i}}: b_{i} \ne \emptyset, 1 \le i \le L\}$
has a permissible updating cycle.
If $\Delta \ne \emptyset$, then $\Delta \subset b_{i}$ for every $i$.
\end{lemma}
\begin{proof}
Without loss of generality,
let $\langle \langle a_1,\ldots, a_L\rangle \rangle$ be a permissible updating cycle.
When $u\in \Delta=(\bigcup_{i=1}^{L} b_i) \backslash (\bigcup_{i=1}^{L} a_i)$,
 $u\in b_j$ for some $j$, but $u\not\in a_i$ for all $i$.
Because of {\sf Rule A}, we have $u\in b_j\subseteq a_{j-1}\cup b_{j-1}$.
Hence, $u$ must also belongs to $b_{j-1}$. By induction, $u$ belongs to every $b_i$,
which implies that $\Delta \subset b_{i}$ for every $i$.
\end{proof}

\begin{alg}[conditioned ICR]\label{alg:3}\rm
%Let the conditional model be  ${\cal A}=\{ f_{a_{i^{\ast}}|b_{i^{\ast}}}: b_{i^{\ast}} \ne \emptyset, 1 \le i \le L\}$.
%When $\Delta \ne \emptyset$ and  $\langle 1^{\ast},\ldots, L^{\ast}\rangle$ is a permissible updating cycle,
%ICR will synthesize from ${\cal A}$ distributions that are conditioned on $\Delta$.
Let $\{ f_{a_{i}|b_{i}}: b_{i} \ne \emptyset, 1 \le i \le L\}$
be a conditional model having a permissible updating cycle.
When $\Delta \ne \emptyset$, ICR will synthesize densities that are conditioned on $\Delta$.
\end{alg}

%\begin{alg}[conditioned ICR]\label{alg:3}\rm
%Let the conditional model be  ${\cal A}=\{ f_{a_{i^{\ast}}|b_{i^{\ast}}}: b_{i^{\ast}} \ne \emptyset, 1 \le i \le L\}$.
%When $\Delta \ne \emptyset$ and  $\langle 1^{\ast},\ldots, L^{\ast}\rangle$ is a permissible updating cycle,
%ICR will synthesize from ${\cal A}$ distributions that are conditioned on $\Delta$.
%\end{alg}

Let $\langle \langle a_1,\ldots, a_L\rangle \rangle$ be a permissible updating cycle.
The initial density is $q^{(1)}_{(a_{1} \cup b_{1} )
\backslash \Delta | \Delta}= f_{a_{1} | b_{1}}
q^{(0)}_{(b_{1} \backslash \Delta) | \Delta}$, where
$q^{(0)}_{(b_{1} \backslash \Delta) | \Delta}$ is any
conditional density of $(b_{1} \backslash \Delta)$ given  $\Delta$.
Every subsequent  distribution produced by ICR is also conditioned on $\Delta$.
A simple example is $\{ f_{1|23}, f_{2|13}\}$.
Another example is $\{f_{12|345}, f_{3|245}\}$ which permits
${\mathbb P}$ mapping from ${\cal C}_3$ onto ${\cal C}_{12}$ conditioned on
$\Delta=\{x_4, x_5\}$, and ${\mathbb P}$ mapping from ${\cal C}_{12}$ back
onto ${\cal C}_{3}$ conditioned on $\Delta$. Using Algorithm~\ref{alg:3},
ICR synthesizes $\pi_{123|45}^{(3,12)}$ and $\pi_{23|45}^{(12,3)}$
from $\{f_{12|345}, f_{3|245}\}$.
In the following, we concentrate on Algorithm~\ref{alg:2},
because most discussions apply to Algorithm~\ref{alg:3} with
additional conditioning on $\Delta$.

\begin{definition}
For CSM: $\{f_{a_i|b_i}: b_i \ne \emptyset, 1 \le i \le L\}$, let
${\cal C}_{a_i}=\{ f_{a_i|b_i}v_{b_i}\}$, and
$\langle \langle a_1,\ldots, a_L\rangle \rangle$ be a permissible updating cycle.
A collection of densities,
$\{\pi^{(a_{i+1},\ldots,a_L,a_1,\ldots,a_i)} \in {\cal C}_{a_i}: 1 \le i \le L\}$, are said to
be mutually stationary when ${\mathbb P}(\pi^{(a_{i+1},\ldots,a_L,a_1,\ldots,a_i)})
= \pi^{(a_{i+2},\ldots,a_L,a_1,\ldots,a_{i+1})}$ for every $i$,
with $(L+1)\equiv 1$.
\end{definition}

Mutually stationary distributions have the following properties:
\begin{enumerate}[(a)]
\item Each set of $\{\pi^{(a_{i+1},\ldots,a_L,a_1,\ldots,a_i)}\}$ is
associated with a specific permissible updating cycles.
\item Every $\pi^{(a_{i+1},\ldots,a_L,a_1,\ldots,a_i)}$ is stationary with
 respect to ${\mathbb P}^{L}$, i.e.,
${\mathbb P}^{L}(\pi^{(a_{i+1},\ldots,a_L,a_1,\ldots,a_i)})=\pi^{(a_{i+1},\ldots,a_L,a_1,\ldots,a_i)}$.
\item For saturated CSM, $\{\pi^{(1,2)},\pi^{(2,1)}\}$,
$\{\pi^{(1,2,3)},\pi^{(2,3,1)},\pi^{(3,1,2)}\}$ and
$\{\pi^{(a_2,\ldots,a_L,a_1)}, \pi^{(a_3,\ldots,a_L,a_1,a_2)},$
$\ldots, \pi^{(a_1,a_2,\ldots,a_L)}\}$
are mutually stationary.
\item Neighboring marginal densities satisfy
$\pi^{(a_{i+1},\ldots,a_L,a_1,\ldots,a_i)}_{b_{i+1}} = \pi^{(a_{i+2},\ldots,a_L,a_1,\ldots,a_{i+1})}_{b_{i+1}}$, i.e., condition~(H2) for every $i$.
\item For a compatible CSM having $\pi^*$ as its joint,
$\{\pi^*_{a_i \cup b_i}: i\le i \le L \}$ satisfy
${\mathbb P} (\pi^*_{a_i \cup b_i})= \pi^*_{a_{i+1} \cup b_{i+1}}$,
hence are mutually stationary.
\item If one $\pi^{(a_{i+1},\ldots,a_L,a_1,\ldots,a_i)}$ is known, the other $L-1$ stationary densities
can be computed via mapping ${\mathbb P}$ cyclically.
For example, when $\pi^{(1,2)}$ is known, $\pi^{(2,1)}= {\mathbb P}(\pi^{(1,2)})$.
\item Only for a saturated CSM, $\{\pi^{(a_{i+1},\ldots,a_L,a_1,\ldots,a_i)}\}$
are all joint densities.
\item The assertion of the existence of
$\{\pi^{(a_{i+1},\ldots,a_L,a_1,\ldots,a_i)}\}$ is always true for
totally positive CSM. Otherwise, the existence depends on whether
$\pi^{(a_{i+1},\ldots,a_L,a_1,\ldots,a_i)}_{b_{i+1}}$ is a bona fide marginal distribution of
$b_{i+1}$ for $1\le i \le L$.
%In rare cases, the marginalized
%support of $\pi^{(a_{i+1},\ldots,a_L,a_1,\ldots,a_i)}$ is not the
%support of $x_{b_{i+1}}$,
%then the existence of mutually stationary distributions could be in question.
\end{enumerate}
Therefore, we first determine a permissible updating cycle,
say $\langle \langle a_1,\ldots,a_L\rangle \rangle$, then ICR will compute
$\{\pi^{(a_{i+1},\ldots,a_L,a_1,\ldots,a_i)}\}$.
In the following proofs, the CSM is $\{f_{a_i|b_i}:1\leq i\leq L\}$,
$c_i=a_i\cup b_i$, symbol $x_{c_i}$ denote values of $(x_j: j \in c_i)$
and ${\cal C}_{a_i}=\{h_{c_i}:h_{a_i|b_i}=f_{a_i|b_i}\}$.

\begin{lemma}\label{lem:shorten}
Assume ${\cal C}_{a_i} \rightharpoonup {\cal C}_{a_j}$ is permissible.
For any two densities $h$ and $g$ in ${\cal C}_{a_i}$,
mapping both by ${\mathbb P}$ onto ${\cal C}_{a_j}$ decreases their K-L divergence.
That is, $I(h;g)>I({\mathbb P}(h);{\mathbb P}(g))$.
\end{lemma}
\begin{proof}
First, we have
\begin{eqnarray*}
\lefteqn{I(h;g)}\\
&=&\sum_{x_{c_i}}h(x_{c_i})\log\frac {h(x_{c_i})}{g(x_{c_i})}\\
&=&\sum_{x_{b_j}}h_{b_j}(x_{b_j})\sum_{x_{{c_i}\backslash b_j}}h_{{c_i}\backslash b_j|b_j}(x_{{c_i}\backslash b_j}|x_{b_j})\left(\log\frac {h_{{c_i}\backslash b_j|b_j}(x_{{c_i}\backslash b_j}|x_{b_j})}{g_{{c_i}\backslash b_j|b_j}(x_{{c_i}\backslash b_j}|x_{b_j})}+\log\frac{h_{b_j}(x_{b_j})}{g_{b_j}(x_{b_j})}\right)\\
&=&\left[\sum_{x_{b_j}}h_{b_j}(x_{b_j})I(h_{{c_i}\backslash b_j|b_j}(x_{{c_i}\backslash b_j} |x_{b_j});g_{{c_i}\backslash b_j|b_j}(x_{{c_i}\backslash b_j} |x_{b_j}))\right] +I(h_{b_j};g_{b_j}).
\end{eqnarray*}

It is easy to see that
$I({\mathbb P}(h);{\mathbb P}(g))=I(h_{b_j};g_{b_j})$, because
${\mathbb P}(h)$ and ${\mathbb P}(g)$ have the same conditional density $f_{a_j|b_j}$.
Hence,
\[
I(h;g)-I({\mathbb P}(h);{\mathbb P}(g))
=\sum_{x_{b_j}}h_{b_j}(x_{b_j})I(h_{{c_i}\backslash b_j|b_j}(x_{{c_i}\backslash b_j} |x_{b_j});g_{{c_i}\backslash b_j|b_j}(x_{{c_i}\backslash b_j} |x_{b_j})),
\]
which is strictly positive, unless $h_{{c_i}\backslash b_j|b_j}=g_{{c_i}\backslash b_j|b_j}$ for every $x_{b_j} \in b_j $.
\end{proof}

The following theorem proves that  the $L$ sequences
of  densities produced by ICR converge respectively to
mutually stationary densities.

\begin{theorem}\label{thm:convergence}
For a permissible updating cycle,
say $\langle \langle a_1,\ldots,a_L\rangle \rangle$, assume the corresponding
$L$ mutually stationary densities $\pi^{(a_{i+1},\ldots,a_L,a_1,\ldots,a_i)}$, $1 \le i \le L$
with $(L+1)\equiv 1$, exist.  For every $1 \le i \le L$,
the sequence of  densities produced by Algorithm~\ref{alg:2}, $\{q^{(kL+i)}\}$,
 converge monotonically to $\pi^{(a_{i+1},\ldots,a_L,a_1,\ldots,a_i)}$ in K-L divergence,
as $k$ tends to $\infty$.
\end{theorem}
\begin{proof}
Due to Lemma~\ref{lem:shorten}, we have, for $1 \le i \le L$,
\begin{eqnarray*}
I\left(\pi^{(a_{i+1},\ldots,a_L,a_1,\ldots,a_i)};q^{(kL+i)}\right)
 & >& I\left({\mathbb P}(\pi^{(a_{i+1},\ldots,a_L,a_1,\ldots,a_i)});{\mathbb P}(q^{(kL+i)})\right)\\
 &= & I\left(\pi^{(a_{i+2},\ldots,a_L,a_1,\ldots,a_{i+1})};q^{(kL+i+1)}\right).
\end{eqnarray*}
After applying ${\mathbb P}$ $L$ times, ICR is back
to ${\cal C}_{a_i}$ with
${\mathbb P}^L (\pi^{(a_{i+1},\ldots,a_L,a_1,\ldots,a_i)})
= \pi^{(a_{i+1},\ldots,a_L,a_1,\ldots,a_i)}$,
and ${\mathbb P}^L (q^{(kL+i)})= q^{((k+1)L+i)}$. Thus,
\begin{eqnarray*}
I\left(\pi^{(a_{i+1},\ldots,a_L,a_1,\ldots,a_i)};q^{(kL+i)}\right) &>&
I\left({\mathbb P}^L (\pi^{(a_{i+1},\ldots,a_L,a_1,\ldots,a_i)});{\mathbb P}^L (q^{(kL+i)})\right)\\
&=&
I\left(\pi^{(a_{i+1},\ldots,a_L,a_1,\ldots,a_i)};q^{((k+1)L+i)} \right).
\end{eqnarray*}
Hence,  $I\left(\pi^{(a_{i+1},\ldots,a_L,a_1,\ldots,a_i)};q^{(kL+i)}\right)$
decreases strictly to zero as $k \rightarrow\infty$.
\end{proof}

Because the decrease is monotonic,  Algorithm~\ref{alg:2} may be stopped at $(k+1)$th
cycle when $I\left( q^{(kL+i)}; {\mathbb P}^L (q^{(kL+i)})\right) < 10^{-10}$ for any $i$.
%Theorem~\ref{thm:convergence}
The following corollary provides  theoretical justifications for  PCGS.

\begin{corollary}
Let $\pi$ be a joint distribution of $\mathbb{X}$ and the CSM be
$\{\pi_{i|c_i}:1\leq i\leq d\}$.
Let $(1^{\ast},\ldots, d^{\ast})$ be a permutation of $(1,\ldots,d)$.
When {\rm (a)} $i^{\ast}\in c_{(i+1)^{\ast}}$
and {\rm (b)} $c_{(i+1)^{\ast}}\backslash \{i^{\ast}\}\subseteq c_{i^{\ast}}$
for every $i^{\ast}$ with $(d+1)^{\ast} \equiv 1^{\ast}$,
Algorithm~\ref{alg:2} will synthesis $\{\pi_{i\cup c_i}\}$ from $\{\pi_{i|c_i}\}$.
Moreover, PCGS updating in the order of
$x_{1^{\ast}}\to x_{2^{\ast}}\to\cdots\to x_{d^{\ast}}\to x_{1^{\ast}}\to\cdots$
preserves  stationarity.
\end{corollary}

Another feature of  Algorithm~\ref{alg:2} is that it  can be applied to subgroups of conditionals
after suitably partitioning the CSM;
the rule is that a permissible updating cycle is identified within each subgroup.
%From each subgroup, ICR  generates a batch of marginals and conditionals.
Depending on the CSM, ICR might
be able to synthesize the outcomes of the subgroups---the many local models---into global
joint distributions of ${\mathbb X}$.  We shall name such an approach ``divide-then-ICR''.
%see Examples~\ref{ex:1}, \ref{ex:3} and \ref{ex:4}.
If we depict a CSM as a directed graph,
GS requires a feedback loop that connects every variable of $\mathbb{X}$.
Therefore, the option of partitioning a CSM into subgroups is not available to GS or PCGS.

\section{Examples}\label{sec:3}
\begin{example}[A simple case for divide-then-ICR]\label{ex:1}\rm
Consider the compatible unsaturated CSM, $\{f_3, f_{1|23}, f_{2|13}\}$, with $f_3(0)=f_3(1)=1/2$. \citet{Kuo2018} showed that none of the  six permutations of $(1,2,3)$ can lead PCGS to  generate samples from the correct joint,  because
there is no permissible updating cycle; though, the model is sufficient.
The joint $\pi$ and its two conditional densities $f_{1|23}$ and $f_{2|13}$ are given as follows; moreover, we add an incompatible $g_{2|13}$ to pair with $f_{1|23}$ for showcasing our compatibility check:
\begin{center}
\begin{tabular}{rrrrrrrrr}
\toprule
$x_1$&0&1&0&1&0&1&0&1\\
$x_2$&0&0&1&1&0&0&1&1\\
$x_3$&0&0&0&0&1&1&1&1\\
\midrule
$\pi_{123}$&1/20&3/20&4/20&2/20&3/20&3/20&3/20&1/20\\
$f_{1|23}$&1/4&3/4&2/3&1/3&1/2&1/2&3/4&1/4\\
$f_{2|13}$&1/5&3/5&4/5&2/5&1/2&3/4&1/2&1/4\\
$g_{2|13}$&3/5&1/7&2/5&6/7&4/5&3/4&1/5&1/4\\
\bottomrule
\end{tabular}
\end{center}

The CSM is first partitioned into  $\{f_{1|23}, f_{2|13}\}$  and $\{f_3\}$. Then $\langle \langle 1,2\rangle \rangle$, and
Algorithm~\ref{alg:3} is applied with $\Delta =\{3\}$.  The initial distribution can be any $q^{(0)}_{2|3}$.  The stationary distributions, $\pi_{12|3}^{(2,1)}$  and $\pi_{12|3}^{(1,2)}$, are computed via the following alternating ${\mathbb P}$ mappings:
\[
{\mathbb P}(q^{(2k)})= q^{(2k+1)}_{12|3}\equiv f_{1|23} q^{(2k)}_{2|3} \mbox{ and }
{\mathbb P}(q^{(2k+1)})=q^{(2k+2)}_{12|3} \equiv f_{2|13} q^{(2k+1)}_{1|3},\ k=0,1,2,\ldots.\]
When $q^{(2t+1)}_{1|3}=q^{(2t)}_{1|3}$  and $q^{(2t+2)}_{2|3}=q^{(2t+1)}_{2|3}$, the iterations reaches stationarity.
Numerically, convergence had occurred after seven cycles because $M(t)=I(q^{(2t)}_{1|3};q^{(2t+1)}_{1|3})+I(q^{(2t+1)}_{2|3};q^{(2t+2)}_{2|3})$ drops from $M(0)= 6.7\times 10^{-2}$ to $M(6)=4.7\times 10^{-11}$.  Hence, we have $q^{(13)}_{12|3} = \pi_{12|3}^{(2,1)}$ and $q^{(14)}_{12|3} =\pi_{12|3}^{(1,2)}$.  To check compatibility, we use $\Pi(t)=I(q^{(2t)}_{12|3};q^{(2t+1)}_{12|3})+ I(q^{(2t+1)}_{12|3};q^{(2t+2)}_{12|3})$; it  drops from $\Pi(0)=2.0\times 10^{-2}$ to $\Pi(6)=5.6\times 10^{-11}$, which implies $\pi^{(2,1)}_{12|3}=\pi^{(1,2)}_{12|3}$ and compatibility.  Furthermore, $\pi^{(2,1)}_{12|3} f_3$ reproduces  $\pi_{123}$.
%The synthesise is written as $\langle \langle \langle \langle 1,2 \rangle \rangle, 3 \rangle \rangle$ with two phases.

Now, consider the incompatible case: $\{f_{1|23}, g_{2|13}\}$.  Because $M(0)=2.7\times 10^{-1}$ drops to $M(7)=2.1\times 10^{-11}$, Algorithm~\ref{alg:3} converges after eight cycles.
But $0.92 <\Pi(t)<0.95$, $0\le t\le 10$ never decreases, hence
the two stationary densities are different, which implies that $f_{1|23}$ and $g_{2|13}$
are not compatible.

Let $x_i$ have $m_i$ categories for $i=1,2,3$.  In order to match the joint and the $x_3$ marginal distributions, the number of unknowns is $m_1 m_3+ m_2 m_3-2$, but the number of equations is $m_1 m_2 m_3+ 2 m_3-3$.  In terms of computational effort, Algorithm~\ref{alg:3} is much simpler than solving over-specified linear equations.
\qed
\end{example}

\begin{example}[Permissible updating cycles of an unsaturated CSM]\label{ex:2}\rm
Consider a hypothetical example that an Asian nation applies to become a permanent member
of the Security Council of United Nations (UN). America's vote is conditioned
on Great Britain and France, but not on Russia and China. So its  conditional distribution
is a non-full conditional.  Assume that France's vote would be
conditioned on the other four nations, so its conditional distribution is a full conditional.
Only the joint distribution can express the  probability that this nation
will not receive a veto.  In {\sf Stage I}, each conditional distribution can be estimated from
this nation's voting history in UN and geopolitics; in {\sf Stage II}
 joints will be synthesized from this unsaturated CSM.
Here, we consider a hypothetical model  whose $f_{i|a_i}$ are derived from a
randomly generated $\pi(x_1,\ldots,x_5)$, hence, compatible:
%Here, we consider the following compatible conditional model:
\[{\cal A}=\{f_{1|2345},f_{2|1345},f_{3|145},f_{4|15},f_{5|1234}\}.\]
There are only two out of $5!=120$ updating cycles that are permissible:
$\langle \langle 5,4,3,2,1\rangle \rangle$ and $\langle \langle 5,1,4,3,2\rangle \rangle$.
Therefore, partition of CSM is not needed.
For $\langle \langle 5,4,3,2,1\rangle \rangle$, one cycle of Algorithm~\ref{alg:2} is as follows:
\[q^{(5t+1)}=f_{5|1234}q_{1234}^{(5t)},\ q_{145}^{(5t+2)}=f_{4|15}q_{15}^{(5t+1)},\
q_{1345}^{(5t+3)}=f_{3|145}q_{145}^{(5t+2)},\]
\[q^{(5t+4)}=f_{2|1345}q_{1345}^{(5t+3)},\ q^{(5t+5)}=f_{1|2345}q_{2345}^{(5t+4)}.\]
Every $I$-projection does two operations: marginalization then multiplication.
For some non-full conditionals, marginalization may not be required.
Among the above five steps, $q^{(5t+2)}_{145}$ and $q^{(5t+3)}_{1345}$ are,
respectively, multiplied directly into
$f_{3|145}$ and $f_{2|1345}$ to form $q_{1345}^{(5t+3)}$ and $q^{(5t+4)}$.
When no marginalization is performed, the $q^{(k)}$
 will not conflict with the conditional models.
%and the test of convergence can skip them.
Stop ICR when $q^{(5t+1)}_5 =q^{(5t)}_5$, $q^{(5t+1)}_{234} =q^{(5t+4)}_{234}$, and  $q^{(5t+4)}_1 =q^{(5t+5)}_1$.
Numerically,  ICR iterations will be stopped at $t$ when
\[M(t)=I(q_5^{(5t)};q_5^{(5t+1)})+I(q_{234}^{(5t+1)};q_{234}^{(5t+4)})+I(q_1^{(5t+4)};q_1^{(5t+5)}) < 10^{-10}.\]
When  compatibility is in question, you compute the following $\Pi(t)$:
\[\Pi(t)=I(q^{(5t)};q^{(5t+1)})+I(q^{(5t+1)};q^{(5t+4)})+I(q^{(5t+4)};q^{(5t+5)}).\]
If it drops to $0$, the CSM is compatible, otherwise, not.
%We list the $M_i(t)$  and $\Pi_i(t)$ for the other permissible updating cycle for reference.
The stopping criterion for the other permissible cycle: $\langle \langle 5,1,4,3,2\rangle \rangle$ is
%\begin{eqnarray*}
%M_{2}(t)&=&I(q_5^{(5t)};q_5^{(5t+1)})+I(q_1^{(5t+1)};q_1^{(5t+2)})+I(q_{234}^{(5t+2)};q_{234}^{(5t+5)}),\\
%\Pi_{2}(t)&=&I(q^{(5t)};q^{(5t+1)})+I(q^{(5t+1)};q^{(5t+2)})+I(q^{(5t+2)};q^{(5t+5)}).
%\end{eqnarray*}
%In practice,  we only need one sequence to confirm or deny compatibility.
\[M(t)=I(q_5^{(5t)};q_5^{(5t+1)})+I(q_1^{(5t+1)};q_1^{(5t+2)})+I(q_{234}^{(5t+2)};q_{234}^{(5t+5)}).\]
For both updating cycles, the randomly generated joint distribution is recovered.
\qed
\end{example}

\begin{example}\label{ex:3}\rm
\citet[Section~4]{Arnold1996} considered the unsaturated CSM:
$\{f_{1|2345},f_{2|345},$ $f_{3|145},f_{4|25},f_{5|13}\}$; they
used a procedure that is equivalent  to recursive factorization to derive the joint density.
We illustrate divide-then-ICR here.  First, divide the CSM into
$ \{f_{1|2345},f_{2|345},f_{3|145}\}, \{f_{4|25} \}$,  and $\{f_{5|13}\}$ because
 $\langle \langle 3,2,1\rangle \rangle$; $\langle \langle 123,4\rangle \rangle$  and
$\langle \langle 1234,5\rangle \rangle$ hold.
\begin{enumerate}
  \item Phase 1: Algorithm~\ref{alg:3} produces $\pi_{123|45}^{(3,2,1)}, \pi_{13|45}^{(2,1,3)}, \pi_{23|45}^{(1,3,2)}$ condition on $\{4,5\}$. To build a joint, only $\pi_{123|45}^{(3,2,1)}$ needs to be used in the next phase.
  \item Phase 2: Algorithm~\ref{alg:3} uses $\{\pi_{123|45}^{(3,2,1)}, f_{4|25}\}$ to build $\pi_{1234|5}^{(4,123)}$  and $\pi_{24|5}^{(123,4)}$ conditioned on $\{5\}$.
  \item Phase 3: Algorithm~\ref{alg:2} uses $\{ \pi_{1234|5}^{(4,123)}, f_{5|13}\}$ to build a joint $\pi_{12345}^{(5,1234)}$ and a marginal $\pi_{135}^{(1234,5)}$.
\end{enumerate}

When the CSM is compatible, $\pi_{12345}^{(5,1234)}$ is the joint producing the CSM.
The synthesis is written as $\langle \langle ~\langle \langle ~\langle \langle 1,2,3\rangle \rangle , 4\rangle \rangle ,5\rangle \rangle$. \qed
\end{example}

\begin{example}[Embedding a CSM  like a jigsaw puzzle]\label{ex:4}\rm
 Let the CSM be $\{f_{2|1}, f_{3|2}, f_{1|3}, f_{4|123},$ $ f_{5|1246}, f_{6|1245}, f_{3^*|12456},$ $ f_{6^*|12345}\}$, where $3^*$ and $6^*$ indicate the variables appear twice in the model.  We divide CSM into $4$ subgroups: $\{f_{2|1}, f_{3|2}, f_{1|3} \}$, $\{ f_{4|123}\}$, $\{ f_{5|1246}, f_{6|1245}\}$, $\{ f_{3^*|12456}, f_{6^*|12345}\}$, and use Algorithm 2 or 3 to consolidate the conditionals in each group into: marginals: $\{\pi_{12}, \pi_{23}, \pi_{13}\},$ and conditionals: $f_{56|124}, f_{36|1245}$, respectively.

 In order to incorporate $f_{4|123}$, we need the marginal $\pi_{123}$ which is missing from the CSM, so the CSM is not sufficient.    Recall the three-way log-linear model:
 \[
 \log \pi_{ijk}= \mu +\mu^1_i + \mu^2_j+ \mu^3_k + \mu^{12}_{ij}+ \mu^{23}_{jk}+\mu^{13}_{ik}+ \mu^{123}_{ijk}.
 \]
 In order to obtain $\pi_{123}$, an assumption about the three-way iterations is required.  Either $\mu^{123}_{ijk}= 0$ or $\mu^{123}_{ijk}= constant$ is most common;   other possibilities may need some subject-matter knowledge.  Once $\mu^{123}$ are settled, use iterative proportional fitting algorithm (IPF) along with $\{\pi_{12}, \pi_{23}, \pi_{13}\}$ to obtain $\pi_{123}$. %, where the $~^#~$ indicates the values can change depending on the assumption about $\mu^{123}$.
 Combining $\pi_{123}$ and $f_{4|123}$ gives $\pi_{1234}$, which will be marginalized into $\pi_{124}$ to be combined with $f_{56|124}$ to form $\pi_{12456}$.  This  distribution can be reduced to $\pi_{1245}$ to be matched with $f_{36|1245}$ to form a joint distribution $\pi_{123456}$.  %Keep in sight that there could be two version of
%$f_{56|124}$ and $f_{36|1245}$ due to different updating sequences. %, while $\pi_{1234}^#$ is unique once $\mu^{123}$ are chosen.
%The synthesis is written as $\langle \langle ~\langle \langle ~\langle \langle ~\langle \langle 1,2,3\rangle \rangle ,4\rangle \rangle ,\langle \langle 5,6\rangle \rangle , \rangle \rangle , \langle \langle 3^*, 6^* \rangle \rangle ~ \rangle \rangle$.
\qed
\end{example}

\begin{example} [A sticky conditional model for GS]\rm
Consider the following compatible conditionals:
\begin{center}
\begin{tabular}{ccccccc}
\toprule
$x_1$&0&1&0&1&0&1\\
$x_2$&0&0&1&1&2&2\\
\midrule
$f_{1|2}$&100000/100001&1/100001&100000/100001&1/100001&7/8&1/8\\
$f_{2|1}$&200000/700007&2/8&500000/700007&5/8&7/700007&1/8\\
\bottomrule
\end{tabular}
\end{center}
, which are derived from the following joint density:
\[\pi=\left(\frac {200000}{700015},\frac {2}{700015},\frac {500000}{700015},\frac {5}{700015},
\frac {7}{700015},\frac {1}{700015}\right).\]
It would be difficult for GS to explore the support because the concentration of probabilities at $(0,0)$  and $(0,1)$. Here we show that ICR will not be hindered by the sticky cells.

For ICR, %updates in the cycle of ${\cal C}_1 \rightharpoonup {\cal C}_2 \rightharpoonup {\cal C}_1$, and
 $M(4)=3.8\times 10^{-11}$ indicates convergence after five rounds of ICR.
The mutual K-L divergence between $\pi^{(1,2)}$ and $\pi^{(2,1)}$ is $\Pi(4)=3.9\times 10^{-11}$, thus confirms that the model is compatible, and $\pi$ is reproduced.
Next, GS is used to produce $5$ batches of size $1,000,000$ samples from $\{f_{1|2},f_{2|1}\}$; the burn-in is set at $100,000$.
Let $g^{(s)}$, $s=1,\ldots,5$, be the empirical pdf with $g^{(1)}$ based on the first
$1,000,000$ samples, and the other $g^{(i)}$s based on $4$ increments of  $1,000,000$ additional samples.
The accuracy of GS  are measured by  discrepancies, $I(g^{(s)};\pi)+I(\pi;g^{(s)}), s=1,\ldots,5$.
Last, let $T_1$ and $T_2$ be the transition matrices based on $f_{1|2}$ and $f_{2|1}$, respectively.
The power method  uses the averages of the six rows of $(T_1T_2)^n$ as the approximations to $\pi$.
Let $p^{(t)}$ be the distribution by power-method approximations, which stops at $t=5$ with  $I(p^{(5)};\pi)+I(\pi;p^{(5)})<10^{-10}$, where $\pi$ is the target joint distribution.

\begin{figure}
\begin{center}
\includegraphics[width=3in]{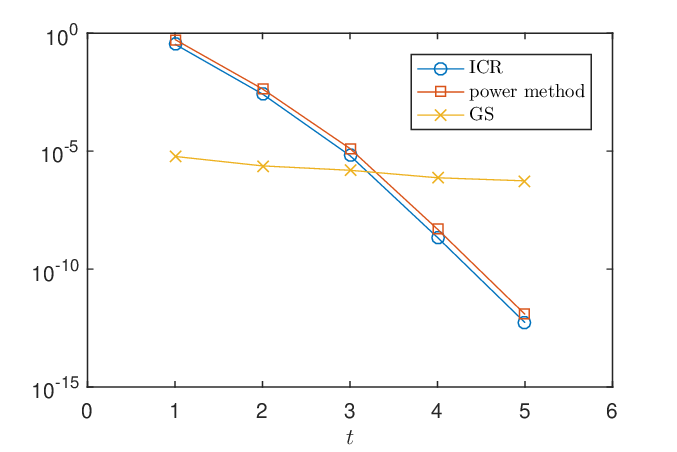}
\end{center}
\vspace{-0.8cm}
\caption{\small Comparisons of the speeds of convergence among ICR ($I(q^{(2t)};\pi)+I(\pi;q^{(2t)})$), the power method ($I(p^{(t)};\pi)+I(\pi;p^{(t)})$), and GS ($I(g^{(t)};\pi)+I(\pi;g^{(t)})$).\label{fig:1}}
\end{figure}

In Figure~\ref{fig:1}, ICR converges a bit faster than the power method, while the additional
$4$ million GS samples shows little improvement.% for the approximations of GS.
In terms of efficiency, CPU times (second) of ICR, power method, and GS
are $0.006$, $0.019$ and $114$, respectively.  The  CPU time consumed by
GS makes it impractical to deal with problems having sticky
issue \citep[p.~354]{Williams2001}, also see \citet[Example~5]{Kuo2018} for a sticky Gaussian model.
Sticky issue slows down sample-based exploitation of the support, but it dose not affect distribution-based ICR or power method.\qed
%But power method is impractical for large $d$, because the dimension of its transition matrix grows exponentially with the number of variables.  %Comparisons between ICR and the power method can be found in Section 4.4. \qed
\end{example}

\begin{example} [Conditional models with disjoint support]\label{ex:5}\rm
Consider a compatible model ${\cal A}_1=\{f_{1|234},f_{2|134},f_{3|124},f_{4|123}\}$ and
an incompatible model ${\cal A}_2=\{f_{1|234},f_{2|134}, f_{3|124}, g_{4|123}\}$, whose
conditional densities are detailed as follows:

\begin{center}
\begin{tabular}{ccccccccc}
\toprule
$x_1$&0&0&1&1&0&0&1&1\\
$x_2$&0&1&0&1&0&1&0&1\\
$x_3$&0&0&1&1&0&0&1&1\\
$x_4$&0&0&0&0&1&1&1&1\\
\midrule
$f_{1|234}$&1&1&1&1&1&1&1&1\\
$f_{2|134}$&1/8&7/8&2/5&3/5&5/12&7/12&1/5&4/5\\
$f_{3|124}$&1&1&1&1&1&1&1&1\\
$f_{4|123}$&1/6&1/2&2/3&3/7&5/6&1/2&1/3&4/7\\
$g_{4|123}$&1/6&3/10&2/3&3/7&5/6&7/10&1/3&4/7\\
\bottomrule
\end{tabular}
\end{center}

Their support $S$ is the union of two disjoint regions
$S_1=\{(0,0,0,0),(0,1,0,0),(0,0,0,1),$ $(0,1,0,1)\}$ and
$S_2=\{(1,0,1,0),(1,1,1,0), (1,0,1,1),(1,1,1,1)\}$.
We will use three different marginal distributions: $u$, $v$ and $w$  to show how
they affect the stationary distributions:
%In order to verify uniqueness, the following three initial densities are chosen.
\[
\begin{tabular}{|c|c|c|c|c|c|c|}
\hline
$x_1$&$x_2$&$x_3$&$x_4$&$u$&$v$&$w$\\
\hline
0&0&0&0&1/8&1/20&1/15\\
\hline
0&1&0&0&1/8&3/20&2/15\\
\hline
0&0&0&1&1/8&2/20&3/15\\
\hline
0&1&0&1&1/8&4/20&4/15\\
\hline
\multicolumn{4}{|c|}{total of $S_1$}&1/2&1/2&2/3\\
\hline\hline
1&0&1&0&1/8&1/10&1/15\\
\hline
1&1&1&0&1/8&1/10&1/15\\
\hline
1&0&1&1&1/8&1/10&1/15\\
\hline
1&1&1&1&1/8&2/10&2/15\\
\hline
\multicolumn{4}{|c|}{total of $S_2$}&1/2&1/2&1/3 .\\
\hline
\end{tabular}\]

Notice that $u$ is  the uniform distribution, and
$\sum_{x\in S_i}v(x)=\sum_{x\in S_i}u(x)\neq \sum_{x\in S_i}w(x)$.
Let  $p^{(0)}=f_{4|123}u_{123}$, $q^{(0)}=f_{4|123}v_{123}$ and $r^{(0)}=f_{4|123}w_{123}$ be the
initial distributions of  ICR, which uses $\langle \langle 1,2,3,4\rangle \rangle$ as the updating cycle. %${\cal C}_1\rightharpoonup {\cal C}_2 \rightharpoonup {\cal C}_3\rightharpoonup {\cal C}_4$.
The three sequences of joints are, respectively,
\[p^{(4k+i)}=f_{i|-i}p_{-i}^{(4k+i-1)},\ q^{(4k+i)}=f_{i|-i}q_{-i}^{(4k+i-1)},\
r^{(4k+i)}=f_{i|-i}r_{-i}^{(4k+i-1)},\]
where $i=1,\ldots,4$.
The convergence of $p^{(k)}$ is determined by
\[M_p(k)=I(p_1^{(4k)};p_1^{(4k+1)})+I(p_2^{(4k+1)};p_2^{(4k+2)})+I(p_3^{(4k+2)};p_3^{(4k+3)})+
I(p_4^{(4k+3)};p_4^{(4k+4)}).\]
We stop  ICR at time $t_p$ when $M_p(t_p)<10^{-10}$.
The $M_q(k)$ and $M_r(k)$ are similarly defined,
so are the stopping times $t_q$ and $t_r$.% for $q^{(t)}$ and $r^{(t)}$ are, respectively,
%determined by similarly defined $M_q(t)$ and $M_r(t)$.
~~Figure~\ref{fig:2} plots $M_p(t)$, $M_q(t)$ and $M_r(t)$ vs. $t$, and they all indicate
fast convergence with $t_p=5$, $t_q=4$ and $t_r=4$, respectively.
After convergence, we obtain three batches of stationary joint distributions:
$\{p^{(4t_p+i)}\}$, $\{q^{(4t_q+i)}\}$, $\{r^{(4t_r+i)}\}$, where $i=1,\ldots,4$ are associated with ${\cal C}_i$.

\begin{figure}
\begin{center}
\includegraphics[width=4in]{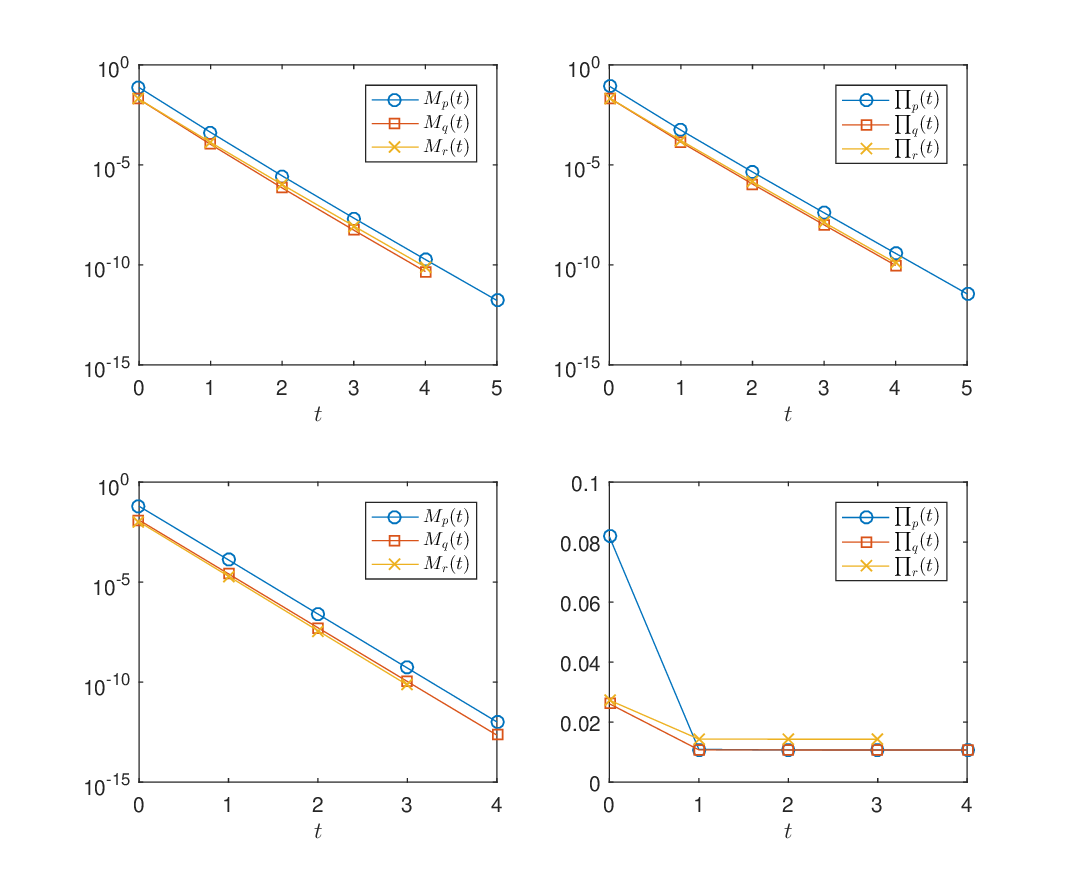}
\end{center}
\vspace{-0.8cm}
\caption{\small The upper panel is for the compatible ${\cal A}_1$, where the left plot is
$M_x(t)$ for detecting the convergence of ICR, and the right plot is $\Pi_x(t)$ indicating compatibility of ${\cal A}_1$,  where $x=p,q,r$;
the lower panel is for incompatible ${\cal A}_2$,  where the convergence indicators, $M_x(t)$, (lower left)  decrease  to $0$, but every $\Pi_x(t)$  indicates multiple stationary distributions (lower right) from ${\cal A}_2$, hence incompatible. \label{fig:2}}
\end{figure}

Compatibility is equivalent to within-group consistency, whose discrepancy is measured by
\[\Pi_p(t)=I(p^{(4t)};p^{(4t+1)})+I(p^{(4t+1)};p^{(4t+2)})+I(p^{(4t+2)};p^{(4t+3)})+I(p^{(4t+3)};p^{(4t+4)}).\]
The resulting $\Pi_p(5)=3.6\times 10^{-12}$, $\Pi_q(4)=9.4\times 10^{-11}$ and $\Pi_r(4)=1.3\times 10^{-10}$
indicate that  ${\cal A}_1$ is a compatible CSM, no matter which initial distribution is used.

Uniqueness of stationary distributions is based on within-${\cal C}_4$ consistency; we need only to compare among $p^{(4t_p+4)}$, $q^{(4t_q+4)}$ and $r^{(4t_r+4)}$:
%are needed due to the within-groups consistency.
\begin{eqnarray*}
I(p^{(4t_p+4)};q^{(4t_p+4)})+I(q^{(4t_q+4)};p^{(4t_p+4)})&=&1.2\times 10^{-12},\\
I(p^{(4t_p+4)};r^{(4t_r+4)})+I(r^{(4t_r+4)};p^{(4t_p+4)})&=&0.1155,\\
I(q^{(4t_q+4)};r^{(4t_r+4)})+I(r^{(4t_r+4)};q^{(4t_q+4)})&=&0.1155.
\end{eqnarray*}
The above  informs us that $p^{(4t_p+4)}=q^{(4t_q+4)}$, $p^{(4t_p+4)} \ne r^{(4t_r+4)}$, and $q^{(4t_q+4)}\ne r^{(4t_r+4)}$.
Therefore, stationary distributions indeed depend of the initial distributions, which is expected for reducible Markov chain.

Next, ICR  with  initial distributions $u$, $v$ and $w$ is applied for ${\cal A}_2$.
The $M_p(t)$, $M_q(t)$ and $M_r(t)$ are plotted against $t$ in lower panel of Figure~\ref{fig:3}.
The left plot  indicates fast convergence also for incompatible CSM, with $M_p(4)=1.0\times 10^{-12}$, $M_q(4)=2.2\times 10^{-13}$, and $M_r(3)=7.2\times 10^{-11}$.
To check compatibility, we calculate $\Pi_x$.  Because $\Pi_p(4)=0.0107$, $\Pi_q(4)=0.0107$ and $\Pi_r(3)=0.0143$,  ${\cal A}_2$ is deemed incompatible.
To see the effect of initial distributions, we compute the following K-L divergences:
\begin{eqnarray*}
I(p^{(20)};q^{(20)})+I(q^{(20)};p^{(20)})&=&1.57\times 10^{-15},\\
I(p^{(20)};r^{(16)})+I(r^{(16)};p^{(20)})&=&0.1155,\\
I(q^{(20)};r^{(16)})+I(r^{(16)};q^{(20)})&=&0.1155.
\end{eqnarray*}
%Again, we obtain $p^{(20)} \approx q^{(20)}$, $p^{(20)} \ne r^{(16)}$, and $q^{(20)}\ne r^{(16)}$.
We see that the difference between $u$ and $v$ does not change their stationary distributions.
In summary, this example shows that
\begin{enumerate}[(a)]
  \item  Convergence of ICR is not affected by the compatibility of the model;
  \item  Compatibility is not affected by the choice of the initial distribution, i.e., our compatibility check is independent of the choice of the initial distribution; and
  \item  It is the probability assigned to each disjoint support, $\mbox{Pr}(S_i)$, not the detailed distribution over $S_i$, that determines the stationary distribution.
\end{enumerate}
When the support is partitioned, $\mbox{Pr}(S_i)$ must be carefully guided by  subject-matter knowledge; the $\mbox{Pr}(S_i)$ may also be adjusted iteratively until the joint distribution is more consistent with data. This flexibility of using initial distribution to fine tune the stationary distribution is not available to irreducible CSM. \qed
\end{example}

\section{Discussions}\label{sec:4}
\subsection{Differences between ICR and GS}\label{sec:4-1}
Consider the saturated CSM $\{f_{i|-i}: 1 \le i \le d\}$; let $T_i$ be the transition matrix of $f_{i|-i}$, and $q$ be a vector representing a joint pdf.
It can be shown that
\[ qT_i=f_{i|-i} q_{-i} \equiv {\mathbb P}(q). \]
That is, $q$ transitioned by $T_i$, $I$-projection of $q$ onto ${\cal C}_i$, and replacing the $(x_i|-x_i)$-conditional of $q$ by $f_{i|-i}$ are the same thing.  But the commonality ends here.
We choose conditional replacement because it is the easiest to modify for non-full conditionals.
%while $I$-projection  leads us to use the K-L divergence, rather than variational norm, to prove the convergence.
Also, {\sf Rule A} of Algorithm~\ref{alg:1} is intuitively necessary because it is the only circumstance under which conditional replacement can be executed.
GS is justified by Markov chain which cannot be applied to incompatible or unsaturated CSM.
A popular remedy is to expand   every non-full conditionals into  a full conditional.
But such a practice may blind GS to use an impermissible updating  cycle, and cause GS to sample from  the  distribution that is not the target.
We show that  identification of the permissible updating cycles is critical for the execution of ICR, while
GS does not need to pay  attention to it, because {\sf Rule A} and {\sf Rule B} are automatically satisfied for full conditionals.

\subsection{Use Gibbs ensemble to find the optimal joint distribution}\label{sec:4-2}
Graphically, a conditional model is depicted by a \emph{cyclic} directed graph with feedback loop.
\citet{Heckerman2000} call such a graphical model a dependency network, and their
objective is to synthesize \emph{one} joint distribution from a saturated CSM derived empirically, and without regard to compatibility.  They used GS based on incompatible full conditionals to synthesize,  and coined the term pseudo-Gibbs sampler (PGS).  They claimed that different updating  cycles of PGS will converge to  nearly identical stationary distributions when the data are large; but, statisticians have refuted such a claim.
For example, \citet[p.~268]{Gelman2001} stated ``the simulations (imputations) never converge to a single
distribution, rather the distribution depends upon the order of the updating and when the updating stopped.''   \citet[p.~257]{Casella1996} also stated ``Gibbs samplers based on a set of densities that are not compatible result in Markov chains that are \emph{null}, that is, they are either null recurrent or transient.''
In fact, \citet[p.~267]{Besag2001} stated that PGS's
``theoretical properties are largely unknown and no doubt considerable caution must be exercised.''
\citet{Heckerman2000} called the stationary distributions of PGS, pseudo-Gibbs distributions (PGD).

According to \citet{Breiman2001}, incompatible CSM faces the multiplicity problem:
there are many different models that have about the same merit.
He suggests that ``aggregating over a large set of competing models can
reduce the nonuniqueness, while improving accuracy.''
In addition, the resulting model ``is also more stable.'' \citep[p.~206]{Breiman2001}
\citet{Chen2013} named
the collection of $d!$ PGDs of a saturated CSM as the Gibbs ensemble, and proposed to use a weighed sum of PGDs as the final model. Building the ensemble requires running $d!$ long chains of Gibbs sampling, which makes the computational burden  heavy, if not impossible for large $d$.   For instance, \citet{Chen2015} used two chains of $1,000,000$ GS samples each to approximate $\pi^{(1,2)}$ and $\pi^{(2,1)}$, even though $\{f_{1|2}, f_{2|1}\}$ are two $2 \times 2$  conditionals.
From $d$ full conditionals, ICR produces $d$ PGDs in one batch, hence,  reduces the computational burden by one order.    \citet{Chen2013} considered only  ensemble for saturated CSM, because PGS cannot sample   unsaturated CSM.   As we have shown, the size of the Gibbs ensemble of an unsaturated CSM is considerably less than $d!$, because only permissible updating cycles need to be entertained.  This understanding makes the computations for unsaturated CSM less prohibitive.
%According to Algorithm~\ref{alg:2}, the size of Gibbs ensemble of an unsaturated CSM can be considerably less than $d!$.
In Example~\ref{ex:2}, $\{f_{1|2345},f_{2|1345},f_{3|145},f_{4|15},f_{5|1234}\}$ have only six stationary distributions
in two batches, not $120$.  Gibbs ensemble  optimizes by computing a weighted mixture of these six distributions.
The deviance of the mixture relative to the CSM is smaller than every individual PGD.  Different deviance measures, such as K-L divergence, Pearson chi-square $X^2$, and Freeman-Turkey $F^2$ have been considered; therefore, the optimal joint will be deviance-dependent.

\subsection{Comparisons between the power method and ICR}\label{sec:4-3}
Back to $\{f_{i|-i}: 1\le i \le d\}$, let $T_i$ be the transition matrix of $f_{i|-i}$, and ${\cal T}=T_1\cdots T_d$.
The power method uses the row average of ${\cal T}^k$ as the stationary distribution for ${\cal T}$.
But in practice, the power method often encounters a sparse ${\cal T}$ of enormous size
when $d$ is large; thus it is not  practical.  ICR computes at least as fast as the power method, and it has the following computational advantages:
\begin{enumerate}[(a)]
\item One cycle of ICR computes $d$ stationary densities, while the power method requires $d$  sequences.  For $d=3$, ICR produces mutually stationary joints: $\pi^{(1,2,3)},\pi^{(2,3,1)},$ and  $\pi^{(3,1,2)}$, whereas the power method needs to  evaluate $3$ separate sequences: $(T_1T_2T_3)^k$, $(T_2T_3T_1)^k$ and $(T_3T_1T_2)^k$ until convergence.
\item The size of $2$-dimensional ${\cal T}$ increases exponentially with $d$, while ICR works with $d$-dimensional arrays.
\item The power method cannot be applied to  unsaturated conditional models because the transition matrices of full and of non-full conditionals have different sizes.
\item When  ${\cal T}$ is a reducible matrix, power method often fails.
\end{enumerate}

\subsection{Method of alternating projections (MAP)}\label{sec:4-4}
\begin{figure}
\centering
\includegraphics[width=4in]{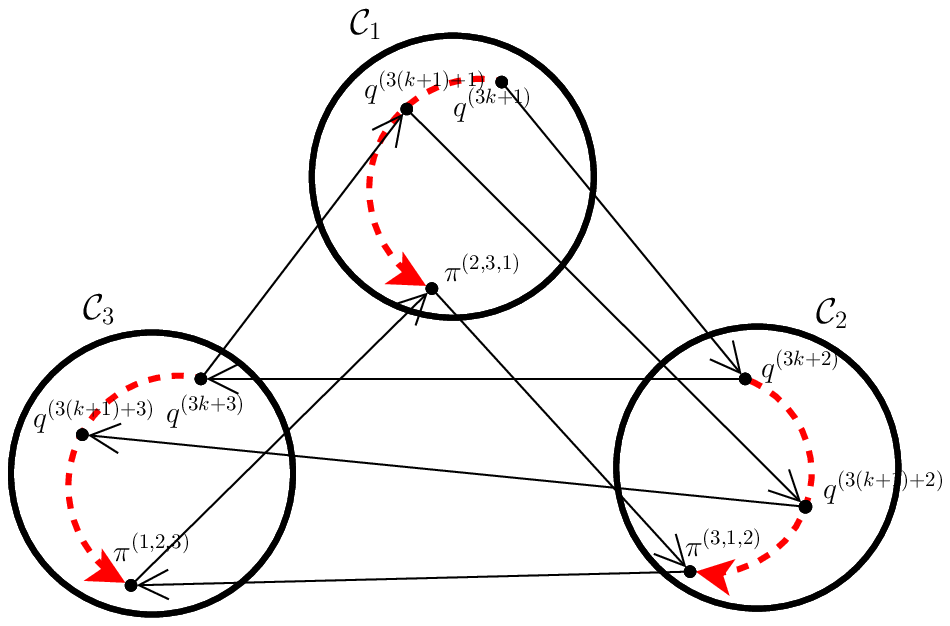}
\caption{\small This figure  illustrates alternating projections for $d=3$ with "$\rightarrow$`` representing ${\mathbb P}$.
Three sequences of joint distributions $q^{(3k+i)}$ within ${\cal C}_i=\{q: q_{i|-i}=f_{i|-i}\}$ converge monotonically to their respective stationary distributions, $\pi^{(j,k,i)}$.  ICR is doing $(T_1 T_2 T_3)^{k} T_1\equiv {\mathbb P}^{(3k+1)}, (T_1 T_2 T_3)^{k} T_1 T_2\equiv {\mathbb P}^{(3k+2)}$ and  $(T_1 T_2 T_3)^{(k+1)}\equiv {\mathbb P}^{(3k+3)}$ in one cycle, where $T_i$ is the transition matrix of $f_{i|-i}$. After convergence, the three $\pi^{i,j,k}$ become mutually stationary.  \label{fig:3}}
\end{figure}

Traditionally, GS considers  ${\cal T}=T_1\cdots T_d$ as one entity; hence,  the effect of individual $T_i$  becomes latent.
However, entertaining $T_i$ separately can gain operational advantage, see, for example,  \citet{Burkholder1961}  and \citet{Burkholder1962}, who used the method of alternating projection (MAP) of \citet{Neumann1950} to find  ``minimal sufficient subfields''.
Also,  it should not be a surprise that our conditional replacement mapping ${\mathbb P}$ onto ${\cal C}_i$ is Burkholder's conditional expectation given ${\cal C}_i$.
More recently, \citet{Diaconis2010} show that the GS is a MAP, when every $T_i$ is considered separately.  When the saturated CSM is compatible,  the proof in \citet{Diaconis2010} guarantees the convergence of ICR in norm, but not in K-L divergence.   However, CSM  often encounter incompatible models having  non-full conditionals. Algorithm~\ref{alg:2} is a MAP, but it is different from ordinary MAP in the following aspects:
\begin{enumerate}[(a)]
\item MAP is commonly used to approximate one fixed point in $\bigcap_i {\cal C}_i$, see \citet{Diaconis2010}.  Here, we show that MAP can also be used to pursue multiple fixed points, one in each ${\cal C}_i$.
\item MAP usually projects onto closed subsets of the same  space, say ${\cal H}=\{$all the joint distributions over ${\cal S}(f_{i|-i}) \}$.   For a saturated CSM,  every ${\cal C}_i$ is a subset of ${\cal H}$.   But the ${\cal C}_i$ defined by a non-full conditional is not a subset of ${\cal H}$, but of a different  space.   Examples here show that MAP  can be applied to closed subsets of different spaces, as long as the projections respect the hierarchy between  spaces, i.e., {\sf Rule A}.
\end{enumerate}
Because of (a) and (b) above, a new concept of  stationarity is needed;
mutual stationarity is better defined collectively, not individually.
Figure~\ref{fig:3} illustrates such pursuits of ${\mathbb P}$ with  $d=3$.  Distributions $q^{(3k+i)}$ within each ${\cal C}_i$ converge monotonically to stationary distribution $\pi^{(j,k,i)}$, and ${\mathbb P}(\pi^{(j,k,i)})= \pi^{(k,i,j)}$.
Minimum context and little background knowledge are required to understand the replacement of conditional distribution, and the simple proof of Theorem~\ref{thm:convergence}.
Our goal is to make ICR, as an algorithm, easily understood  and appreciated by statisticians and data scientists, who have little familiarity with Markov chain theory or Hilbert space.
Another popular MAP algorithm is IPF, which hardly refer to Hilbert space, orthogonal projection or conditional expectation; instead, it is described as replacing marginal densities iteratively, see \citet{Darroch1972} and \citet{Wang1993}.

Finally, much of MAP has been dealing with continuous functions over convex domains.   The algorithm, ``divide-then-ICR,'' and the proof of Theorem~\ref{thm:convergence} can be easily carried over to continuous distributions provided the integrals are finite.   Marginalization of a continuous density is the computatonal obstacle of ICR.  \citet{Cramer1998} studied alternating $I$-projection of a regular Gaussian distribution onto the  intersection of spaces characterized by Gaussian conditionals (a ${\cal C}_i$ defined by a full conditional) and Gaussian marginal distribution (another ${\cal C}_j$ defined by a non-full conditional).  For Gaussian distributions, marginalization is straightforward.  Part of his algorithm \citep[Eq.~2.3]{Cramer1998}
is similar to ICR.  His model placed restrictions on the conditionals that guarantee compatibility (${\cal C}_i \cap {\cal C}_j \ne \emptyset$), hence, has unique stationarity; he did not consider incompatible cases or discrete densities.

\section{Conclusion}\label{sec:5}
When the number of variables is large and the  data size is relatively small, subjective or objective variable selection is necessary, hence, unsaturated conditional models are inevitable.  However,
in the past, only saturated conditional models had been considered---\citet{Besag1974}, \citet{Diaconis2010}, \citet{Kaiser2000}, \citet{Heckerman2000}, \citet{Wang2008}, \citet{Chen2013} and \citet{Kuo2019}---due to lack of computational tools.
On the other front, \citet{Arnold2002,Arnold2004} used linear equations/algebra to check compatibility;  their methods quickly reach the curse of dimension.
ICR is invented to fit unsaturated conditional models, and to check their compatibility using computing, rather than algebra.
ICR provides the channel to apply computing power to solve issues of conditional modeling.
It seems to us that ICR is the right choice for CSM because it is  multiplying by the transition matrix (see Section~ \ref{sec:4-1}), doing $I$-projection (see Section~\ref{sec:2-1}), and performing conditional expectation (see Section~\ref{sec:4-4}), at the same time.
%Moreover, Theorem~\ref{thm:convergence} directly  proves the convergence of PCGS \citep{vanDyk2008}, and Algorithm~\ref{alg:2} shows how to choose valid updating cycles for PCGS.

ICR, along with ``divide-then-ICR'' and parallelization, can efficiently compute all of the mutually stationary distributions, which are called the Gibbs ensemble.
We are in agreement with \citet{Breiman2001} and \citet{Chen2013} that a fair-minded mixture of the Gibbs ensemble is a sensible approach in {\sf Stage III} to resolve the multiplicity problem.
Any practical algorithm must be  easy to scale and  requires little expertise to tune. ICR and  the ensemble optimization  meet both criteria.

\section*{Appendix}
\renewcommand{\thesubsection}{\Alph{subsection}}
\setcounter{subsection}{0}
\subsection{The proof of Pythagoras equality}\label{app1}
Because $\tau \in {\cal C}_1$, it can be written as $\tau=f_{1|2}\tau_2$, and
the K-L divergence between $q$ and $\tau$ is
\begin{eqnarray*}
I(q;\tau)
&=&\sum_{i,j} q(i,j) \log \frac {q(i,j)}
{f_{1|2}(i|j)\tau_2(j)}\\
&=&\sum_{i,j} q(i,j) \log \frac {q(i,j)}
{f_{1|2}(i|j)q_2(j)}+
\sum_{i,j} q(i,j) \log \frac {f_{1|2}(i|j)q_2(j)}{f_{1|2}(i|j)\tau_2(j)}\\
&=& I(q; f_{1|2}q_2) + I(q_2;\tau_2)=I(q; f_{1|2}q_2) + I(f_{1|2}q_2;\tau),
\end{eqnarray*}
because of
\begin{eqnarray*}
I(q_2;\tau_2)
&=& \sum_i f_{1|2}(i|j) \sum_j q_2(j) \log \frac{q_2(j)}{\tau_2(j)}\\
&=& \sum_{i,j}f_{1|2}(i|j)q_2(j) \log \frac{q_2(j)}{\tau_2(j)} \\
&=& \sum_{i,j}f_{1|2}(i|j)q_2(j) \log \frac{f_{1|2}(i|j)q_2(j)}{f_{1|2}(i|j)\tau_2(j)}=I(f_{1|2}q_2;\tau).
\end{eqnarray*}

\end{document}